\documentclass[twocolumn,aps,nopacs,preprintnumbers,nofootinbib, superscriptaddress, amsmath,amssymb]{revtex4}

\usepackage{stmaryrd,tensind} 
\usepackage{mathtools} 
\usepackage{stmaryrd} 

\usepackage[a4paper, total={7in, 10in}]{geometry}
\newcommand{\comm}[1]{}

\usepackage{amsmath}
\usepackage{amsfonts}

\usepackage{amssymb}
\usepackage{amssymb}

\begin{document}

\author{Andre G. Campos}
\email{agontijo@mpi-hd.mpg.de}
\affiliation{Max Planck Institute for Nuclear Physics, Heidelberg 69117, Germany} 

\author{Renan Cabrera}
\affiliation{Arctan, Inc., Arlington, VA 22201, USA} 



\title{ Non-dispersive analytical solutions to the Dirac equation }

\date{\today}

\begin{abstract}

  This paper presents new analytic solutions to the Dirac equation employing a recently introduced method that is based on the
  formulation of spinorial fields and their driving electromagnetic fields in terms of geometric algebras.
  A first family of solutions describe the shape-preserving translation of a wavepacket along \textit{any desired trajectory}
  in the  $x-y$ plane. In particular, we show that the dispersionless motion of a Gaussian wavepacket along both elliptical
  and circular paths can be achieved with rather simple electromagnetic field configurations.
  A second family of solutions involves a plane electromagnetic wave and a combination of generally
  inhomogeneous electric and magnetic fields.  The novel analytical solutions of the Dirac
  equation given here provide important insights into the connection between the quantum relativistic dynamics of electrons
  and the underlying geometry of the Lorentz group.    
\end{abstract} 

 \maketitle

\section{Introduction}

In this work we further expand upon the recently developed framework of \emph{Relativistic Dynamical Inversion} (RDI) \cite{campos2017analytic} 
whose purpose is to solve the 
following problem: Given an arbitrary (desired) spinorial spacetime wavepacket $\psi$, 
find an electromagnetic vector potential  $A_{\mu}$ such that the Dirac equation is satisfied.
This is accomplished by RDI in two steps: First, we verify the attainability of the given evolution 
$\psi$ by assessing the existence of the underlying $A_{\mu}$ leading to valid Maxwell equations (for a proof of this statement see section A of the appendix). 
Second, if it exists, an explicit form of $A_{\mu}$ is obtained which satisfies the Dirac equation for the given $\psi$.
Moreover, the method can also be used to assess for attainable dynamics.

The task of constructing control fields yielding a desired dynamics at all times 
and positions is one of the most important and challenging problems in quantum control.  
In particular, transporting coherent wavepackets without disturbance is a required 
building block in quantum technologies. By breaking down the spinor as a series of local Lorentz transformations
(i.e., Lorentz transformations whose parameters are functions of space and time),
RDI allows for finding analytic solutions which are not feasible by other current methods.
 
Exact solutions of the Dirac equation, a system of four partial differential equations,
are rare. The vast majority of them are for highly symmetric
stationary systems \cite{thaller2013dirac,bagrov2014dirac,eleuch2012analytical}. 
Only a handful of solutions for time dependent dynamics exists 
\cite{wolkow1935klasse,redmond1965solution,bagrov1975solutions,bergou1983relativistic,varro2013new,bialynicki2004particle,Oertel-DiracInversion-2015,kaminer2015self,hayrapetyan2014interaction,Birula-DiracAngularMomentum2017,BarnetRelativisticVortices2017,HeinzlClassicalQuantum2017}.  
For instance, it was long after the first exact time dependent solution was reported by Volkov \cite{wolkow1935klasse}, that its generalization was proposed in \cite{redmond1965solution},
followed by a slightly more general exact solution \cite{bagrov1975solutions}.
Most of the investigations call for either semi-classical methods \cite{lazur2005wkb,popov2006relativistic,popov2005imaginary}
or numerical calculations 
\cite{braun1999numerical,mocken2008fft,bauke2011accelerating,fillion2012numerical,fillion2016galerkin,lv2016numerical,
cabrera2016dirac}. 
In addition to being computationally demanding, commonly used numerical schemes are plagued 
by unphysical artefacts at the fundamental level \cite{hammer2014staggered,hammer2014single}; 
thus, there is a need for  systematic construction of analytic solutions. 
RDI fulfils all these needs by providing stationary 
as well as time-dependent exact solutions in two and three dimensions (see \cite{campos2017analytic} for other solutions). 

Here, RDI is used to construct electromagnetic fields that move a given Dirac spinor along \textit{any desired trajectory}
in the $x-y$ plane without spreading.
In addition, general solutions for a combination of plane electromagnetic waves and electric and magnetic fields along the wave's propagation direction
with arbitrary perpendicular profiles are also constructed.
Illustrations are given for the particular examples of a Gaussian wavepacket moving along both an ellipse and a circle in the $x-y$ plane.
Moreover, we give solutions for a Dirac electron in a combination of a plane electromagnetic wave with a constant and homogeneous
magnetic field along the $z$ axis (known as the Redmond solution \cite{redmond1965solution}) as well as in a combination of a plane electromagnetic wave 
with a constant and homogeneous magnetic field and an electric field of general profile, both along the $z$ axis (first reported by Bagrov et. al. in \cite{bagrov1975solutions}). Our solution generalizes the solutions given by Redmond and Bagrov et. al. in that it also allows for inhomogeneous magnetic fields along the $z$ axis with an arbitrary perpendicular profile.
The analytical solutions of the Dirac equation given here provide important insights into the relativistic dynamics of electrons. 

\section{Methodology of Relativistic Dynamical Inversion.}
The Dirac equation is commonly expressed as 
\begin{align}
\label{Dirac-Equation}
    \gamma^{\mu}[ i c \hbar \partial_{\mu} - c e A_{\mu}] \psi = mc^2 \psi,  
\end{align}
where the summation over repeated indices is adopted, $\psi$ is a four-component complex spinor, 
$m$ 
is the mass, $c$ is the speed of light, $\gamma^{\mu}$ are 
the $4\times 4$ so-called gamma matrices, $A_{\mu}$ is the four-vector potential and  $\mu=0,1,2,3$.

The Dirac equation (\ref{Dirac-Equation}) can be viewed as a ``first quantization'' approximation
to QED. The solutions of Eq. (\ref{Dirac-Equation}) exclude effects such as radiation reaction
and particle creation/annihilation prominent at ultra-relativistic energies. 
Nevertheless, Eq. (\ref{Dirac-Equation}) provides a mean-field description 
of relativistic effects at low and moderate energies. 
A moving Dirac electron generates the current 
$J_D^{\mu}  = \psi^{\dagger} \gamma^0 \gamma^{\mu} \psi$ 
that emits secondary radiation, which is not accounted for by Eq. (\ref{Dirac-Equation}). 
Therefore, a solution of the Dirac equation is physical 
if the energy loss due to the secondary radiation is much smaller than the electron kinetic energy.
This criterion should be satisfied in the applications of the Dirac equation considered in this work.

The Eq. (\ref{Dirac-Equation}) can be written in different forms
emphasizing the geometry of the Lorentz group
\cite{hestenes1967real,hestenes1973local,hestenes1975observables,BaylisBook1996}.
Here, we employ the Hestenes formulation \cite{hestenes1967real} 
where the 
state $\psi$
in Eq. (\ref{Dirac-Equation}) is represented by the matrix $\Psi$,  
\begin{align} \notag
\psi = \begin{pmatrix}
        \psi_1 \\ \psi_2 \\ \psi_3 \\ \psi_4
       \end{pmatrix}
\Longleftrightarrow
\Psi=\begin{pmatrix}
    \psi_{1} & - \psi_2^{*} & \psi_3   & \psi_4^{*} \\
    \psi_2   &   \psi_1^{*} & \psi_4   & -\psi_3^* \\
    \psi_3   &   \psi_4^*  & \psi_1   & -\psi_2^* \\
    \psi_4   &  -\psi_3^*   & \psi_2   &  \psi_1^* 
    \end{pmatrix}.
\end{align} 
obeying the Dirac equation in the matrix form 
\begin{align}  \label{DH-eq}
  \left(\hbar c \partial\!\!\!/ \Psi  \gamma_2\gamma_1 - c e A\!\!\!/ \Psi\right)  =  m c^2 \Psi\gamma_0,
\end{align}
where the Feynman slash notation was employed $A\!\!\!/ = A^{\mu}  \gamma_{\mu}$, $\partial\!\!\!/ = \gamma^{\mu} \partial_{\mu} $
($\mu,\nu = 0,1,2,3$). Note that the matrix $A\!\!\!/$ must have real coefficients $A_\mu$. 
The vector potential may be expressed as a function of the state
\begin{align}
\label{A-inverted}
     e A\!\!\!/&= \hbar \partial\!\!\!/ \Psi  \gamma^2\gamma^1\Psi^{-1} -  m c \Psi \gamma^0\Psi^{-1},
\end{align} 
where
$$
\Psi^{-1}=\frac{\tilde{\Psi}}{\Psi\tilde{\Psi}},\quad \tilde{\Psi}=\gamma_0\Psi^\dagger\gamma_0.
$$
A crucial insight is the spinor factorization for electrons/positrons: 
$\Psi =  \sqrt{\rho} L$, where $\rho$ is a non-negative scalar function modulating the probability density
and $L$ is an invertible matrix representing a  
Lorentz group element \cite{hestenes1967real,hestenes1973local,hestenes1975observables}. 
It is very important to note that for all cases other than a free electron, 
the scalar density $\sqrt{\rho}$ acts as an envelop function
ensuring that the electron's probability distribution $\psi^\dagger\psi$ is normalizable.
Thus, it is always written in the form $\exp(-f(x,y,z,t))$, where $f(x,y,z,t)$ is semi-positive definite. For the
particular case of a free particle, we have $\sqrt{\rho}=1$.

Considering that  $L$ is a member of the special Lorentz group \cite{hestenes1967real,hestenes1973local,
hestenes1975observables}, a spinor $\Psi$  can always be written as the product of
spatial rotations $R$, boosts $\mathcal{B}$  and a transformation
of internal degrees of freedom parametrized by the Yvon-Takabayashi 
angle $\beta$ \cite{yvon1940equations,takabayasi1957relativistic}. Thus, the most general parameterization of the
matrix spinor is \cite{hestenes1967real,hestenes1973local,hestenes1975observables,
BaylisBook1996}
\begin{align}
\label{Psi-factorized}
 \Psi = \sqrt{\rho} \, \mathcal{B} R  e^{\boldsymbol{i}\beta/2},\quad  \boldsymbol{i}=\gamma_0\gamma_1\gamma_2\gamma_3.
\end{align}
The boost $\mathcal{B}$ is written in terms of the velocity components $c\mathbf{v}=c(v^1,v^2,v^3)$ 
(bold symbols denote three dimensional vectors throughout)
\begin{align} \label{boost-u}
 \mathcal{B} =\mathcal{B}(\mathbf{v}) = \frac{  v^{\mu} \alpha_{\mu} +  \mathbf{1} }{ 
 \sqrt{2(1+v^0)} },
\end{align}
with $ v^0 =  \sqrt{1 + \mathbf{v}^2  } $, $\alpha_0$ is the $4\times4$ identity matrix and $\alpha_k=\gamma_k\gamma_0$ are the well known gamma
Dirac matrices; whereas, the spatial rotations are 
parametrized by the angles $   \boldsymbol{\theta} = ( \theta^1 , \theta^2, \theta^3 )$
\begin{align}
\label{R-rotation}
 R =  R( \boldsymbol{\theta} ) = \exp \left(  -\boldsymbol{i} \theta^k \alpha_k/2  \right).
\end{align}
Note that the density $\rho$, velocity $\mathbf{v}$, 
rotation angle $\boldsymbol{\theta}$ and Yvon-Takabayashi angle $\beta$ are in general 
functions of space and time. In this case, we say that the Lorentz transformations encoded in the spinor $\Psi$ are \textit{local}.
Moreover, It must be stressed that \textit{all solutions to the
Dirac equation can be put in the form given by Eq. (\ref{Psi-factorized})}.

RDI is performed in the following way:
Spacetime functions $\rho$, $\mathbf{v}$, $\boldsymbol{\theta}$ and $\beta$
are initially selected to describe a desired dynamics of the Dirac state $\Psi$.
The constructed factorization (\ref{Psi-factorized}) is substituted in Eq. (\ref{A-inverted})
to obtain the vector potential in the matrix form $A\!\!\!/$.  

If the $A_\mu$ are not real,
the proposed dynamics is not reachable with physical fields, and the 
parametrization  $\rho$, $\mathbf{v}$, $\boldsymbol{\theta}$, and $\beta$ 
needs to be modified.

If the $A_\mu$ are real, then we perform the final step of the procedure, consisting in the substitution of both the vector potential
$A_\mu$ and the Dirac spinor describing it, which is simply the leftmost column of $\Psi$, into the Dirac equation (\ref{Dirac-Equation}).
If the Dirac equation is satisfied exactly, then the procedure is completed: The obtained vector 
potential $A_{\mu} = {\rm Tr}\,(A\!\!\!/\gamma_{\mu})/4$ enables to recover the electromagnetic 
fields $F^{\mu \nu} = c\left( \partial^{\mu} A^{\nu} - \partial^{\nu} A^{\mu} \right)$ 
and the source $J^{\nu} = \partial_{\mu} F^{\mu \nu} / (\varepsilon_0 c)$ generating them.  
Provided the current $J^{\nu}$, the obtained fields $F^{\mu \nu}$ necessarily satisfy Maxwell's equations. 
Note that $J^{\nu}$ 
differs from the  current $J_D^{\mu} = {\rm Tr}\,(\Psi \gamma^{\mu}\tilde{\Psi})/4 = \psi^{\dagger} \gamma^0 \gamma^{\mu} \psi$ 
emanating from the Dirac electron. 

Before proceeding to the discussion of the newly found solutions, let us better illustrate the philosophy of RDI by analyzing the case of a free electron in the Hestenes formalism (see \cite{hestenes1967real} for more details). In this case, the Dirac-Hestenes equation (\ref{DH-eq})  becomes
\begin{align}  \label{freeHestenes}
   \hbar c \partial\!\!\!/ \Psi  \gamma_2\gamma_1   =  m c^2 \Psi\gamma_0. 
\end{align}
The two positive energy solutions are
\begin{align}\label{PositivePlaneW}
\Psi^+_i=U_ie^{\gamma_2\gamma_1 p_\mu x^\mu/\hbar},
\end{align}
where the $U_i$ are constant spinorial Lorentz transformations. It is noteworthy that since $\gamma_2\gamma_1=\boldsymbol{i}\alpha_3$, the exponential term is a rotation around the $\gamma_2\gamma_1$ axis by an angle $p_\mu x^\mu/\hbar$. In our examples it is shown that gauge transformations can also be described as rotations around the same axis. Inserting (\ref{PositivePlaneW}) into (\ref{freeHestenes}) gives
$$
p\!\!\!/U_i=mcU_i\gamma_0\rightarrow p\!\!\!/=mcU_i\gamma_0\tilde{U}_i=mcv\!\!\!/,
$$
since $U_i\tilde{U}_i=\boldsymbol{1}$. The Lorentz transformations $U_1$ and $U_2$ correspond to states with spin up and spin down, respectively. They
are explicitly given by
$$
U_1=\mathcal{B}(\mathbf{v}),\quad U_2=\mathcal{B}(\mathbf{v})e^{-\boldsymbol{i}\alpha_2\pi/2}.
$$
The two negative energy solutions are
\begin{align}\label{NegativePlaneW}
\Psi^-_i=V_ie^{\gamma_2\gamma_1 p_\mu x^\mu/\hbar},
\end{align}
where $V_i=\mathcal{B}(-\mathbf{v}) e^{\boldsymbol{i}\pi/2}$, because in this case the Yvon-Takabayashi angle is $\beta=\pi$. Note that $V_i\tilde{V}_i=e^{\boldsymbol{i}\pi}$ although we still have $V_i\gamma_0\tilde{V}_i=v\!\!\!/$. Thus the four momentum becomes
$$
p\!\!\!/=mcv\!\!\!/e^{-\boldsymbol{i}\pi}=-mcv\!\!\!/.
$$
Since $v^0$ is positive, the energy $cp^0$ is negative. 

From the above discussion it becomes clear that whenever we parameterize the spinor by local Lorentz transformations, the addition of the vector potential becomes necessary if the Dirac equation is to be satisfied, in much the same way as when local gauge transformations are performed. Thus, we can claim that all the information about the dynamics of electrons interacting with external fields are contained in the parameterization of the spinor. This aspect of RDI will become
more apparent during the discussion of our novel solutions.


\section{A General solution for motion confined to the x-y plane}
We start with the following Dirac spinor describing an electron wavepacket with spin down, which is in a ground state of some potential,
having zero average velocity at time $t=0$ in the laboratory frame
\begin{align}\label{spinorHomoB}
\psi=e^{-\frac{eB G(x,y) }{4 c \hbar }}\begin{pmatrix}0
   \\ \mathcal{N}(mc^2+\epsilon)\\ 0\\ 0\end{pmatrix},
\end{align}
corresponding to the matrix spinor 
\begin{align}\label{MatrixHomo}
\Psi=e^{-\frac{eB G(x,y) }{4 c \hbar }}\mathcal{N}(mc^2+\epsilon)\begin{pmatrix}
    0& -1 &0   &0 \\
   1 &   0 & 0   & 0 \\
    0   &   0  &0   & -1 \\
    0   &  0  & 1  & 0
    \end{pmatrix} ,
 \end{align}
 where $G$ is a positive real function to be determined later, $\mathcal{N}$ is a normalization constant and $eB>0$ is also a constant.
 Such initial state will always lead to a magnetic field along the $z$ axis in the laboratory frame that is parallel to the $z$-component of the electron's spin.
 Moreover, a peculiar feature of electrons with g-factor $g=2$ is that the ground state energy in this case is exactly $\epsilon=mc^2$ for the case of a
 constant and homogeneous field \cite{thaller2013dirac,johnson1949motion}. 

 We should point out that Eq. (\ref{MatrixHomo}) is of the form (\ref{Psi-factorized}). For instance, since the electron has zero average velocity
 the boost $\mathcal{B}(0)=\boldsymbol{1}$ is simply given by the identity matrix. Moreover, given that the spin of the electron is down (i.e., along the $-\hat{z}$ axis),
 we know from the previous section that
$$
\begin{pmatrix}
    0& -1 &0   &0 \\
   1 &   0 & 0   & 0 \\
    0   &   0  &0   & -1 \\
    0   &  0  & 1  & 0
    \end{pmatrix}=\exp\left(-\frac{\boldsymbol{i}\alpha_2\pi}{2}\right)=R.
$$
However, the scalar density instead of being equal to one as in the free particle case,
it is now given by $\sqrt{\rho}= e^{-\frac{eB G(x,y) }{4 c \hbar }}\mathcal{N}(\epsilon+mc^2)$. Such modification of the matrix spinor necessarily leads to the addition in the Dirac equation
of the following vector potential as can be derived from Eq. (\ref{A-inverted})
\begin{align}
eA^0&=0,\nonumber\\
eA^k&=-\frac{\hbar}{2}\frac{v^0s^3}{\rho}\varepsilon^{kl3}\frac{\partial}{\partial x^l}\rho,
\end{align}
with $v^0=-s^3=1$ since
\begin{align}\label{spinVector}
\rho v\!\!\!/&=\Psi\gamma_0\tilde{\Psi}=e^{-\frac{eB G(x,y) }{2 c \hbar }}\mathcal{N}^2(\epsilon+mc^2)^2\gamma_0,\nonumber\\
\rho s\!\!\!/&=\Psi\gamma_3\tilde{\Psi}=-e^{-\frac{eB G(x,y) }{2 c \hbar }}\mathcal{N}^2(\epsilon+mc^2)^2\gamma_3.
\end{align}
Hence, this confirms our claim that the particular spinor parametrization (\ref{MatrixHomo}) will always lead to some
type of magnetic field along the $\hat{z}$ axis. In the case that $G(x,y)=c(x^2+y^2)$ we recover the ground state of an electron in a constant
and homogeneous magnetic field $e\boldsymbol{B}=\{0,0,-eB\}$. Other choices for the free function $G$ will generally lead to inhomogeneous magnetic fields.

The goal is to translate the electron along an arbitrary trajectory in the $x-y$ plane. In order to do so, we apply to the matrix spinor (\ref{MatrixHomo})
a boost with velocity $f'(t)$ along $x$ and $g'(t)$ along $y$
\begin{align}\label{Boost}
\Psi_b=\mathcal{B}(\mathbf{v})\Psi,\quad
\mathbf{v}=\left\{\gamma \frac{f'(t)}{c},\gamma \frac{g'(t)}{c},0\right\},\nonumber\\
\end{align}
where
$$
\gamma=\frac{c}{\sqrt{c^2-f'(t)^2-g'(t)^2}},
$$
along with the following transformations also performed on the initial state $\Psi$
\begin{align*}
 &x'= x-f(t) \\
 &y'= y-g(t) \\
 &\rho'=\rho(x',y')/\gamma.
\end{align*}
The boosted Dirac spinor $\psi_b$ extract from $\Psi_b$ is then
\begin{align}\label{boostedPsi}
&\psi_b=\sqrt{\frac{\rho'}{2}}\left(\begin{array}{cc}
0\\
   \sqrt{1+\gamma}\\
    \frac{\gamma \left(f'(t)-i
   g'(t)\right)}{
   c\sqrt{1+\gamma}}\\
  0\\
   \end{array}\right),\nonumber\\
 &\sqrt{\rho'}=c^{3/2} m \sqrt[4]{c^2- \left(f'(t)^2+g'(t)^2\right)}\, e^{-\frac{eB
   G\left(x',y'\right)}{4 c \hbar }} \nonumber\\
  & \times \mathcal{N}(\epsilon+mc^2).
\end{align}
The electron's velocity becomes
\begin{align}
v\!\!\!/&=\gamma\left(\gamma_0+\gamma_1\frac{f'(t)}{c}+\gamma_2\frac{g'(t)}{c}\right),\nonumber\\
\end{align}
Note that the spin vector continue to have the form (\ref{spinVector}).
From Eq. (\ref{A-inverted}) we get the following components of the vector potential
\begin{align}\label{VecPot}
eA^0&=\frac{\hbar}{2}\left(\left[\frac{1-\gamma}{c}\right]\frac{d}{dt}\arctan\left(\frac{g'(t)}{f'(t)}\right)+(\boldsymbol{s}\times\boldsymbol{v})\cdot\vec{\nabla}\ln\rho\right)\nonumber\\
&-mc\gamma,\nonumber\\
eA^1&=-\frac{\hbar}{2\rho}\left(\gamma\frac{\partial\rho}{\partial y}+\frac{\partial}{c\partial t}(\rho v^2)\right)-mcv^1,\nonumber\\
eA^2&=\frac{\hbar}{2\rho}\left(\gamma\frac{\partial\rho}{\partial x}+\frac{\partial}{c\partial t}(\rho v^1)\right)-mcv^2,\nonumber\\
eA^3&=0.
\end{align}
The solution we found is a generalization of the 2D solutions found in Ref. \cite{campos2017analytic}. The electron's trajectory is given by the free real functions $f(t)$ and $g(t)$. In the
appendix we prove that the spinor (\ref{boostedPsi}) exactly satisfies the Dirac equation with the vector potential (\ref{VecPot}).

Before proceeding, let us give a more intuitive explanation of the spinor parametrization (\ref{Boost}). In general, the matrix spinor $\Psi$ is an active Lorentz transformation
describing the motion of the electron as seen by an observer in the laboratory frame. Thus, the \textit{local} Lorentz boost $\mathcal{B}$ in Eq. (\ref{Boost}) simply means that
the observer in the laboratory frame sees the electron moving with a varying velocity. The observer thus concludes that the electron is being acted on by a force.
Since what we can measure are trajectories and not fields, we infer from the spinor parameterization the electromagnetic fields causing the observed motion of the electron.
This feature is at the core of RDI, thus being a crucial property of all the solutions discussed in this work.

\subsection{Gaussian tracing out an ellipse without dispersion.}
As an illustration of the newly found solution, we now consider an electromagnetic field that moves a Gaussian wave packet along an ellipse in the $x-y$ plane without distortion.
We choose the following functions
\begin{align*}
&f(t)=a_1\cos(\omega t),\quad g(t)=a_2\sin(\omega t),\nonumber\\
&G(x,y)=c(x^2+y^2),
\end{align*}
where $a_1,a_2$ are the semi-axes of the ellipse. The vector potential is calculated from Eqs. (\ref{VecPot}) for the given functions. 

According to RDI, the electromagnetic fields generating the dynamics
consists of a time dependent homogeneous magnetic field $\boldsymbol{B}$ perpendicular to a planar electric field 
which co-rotates in the $x-y$ plane with the electron. 

\begin{figure}
  \centering
      \includegraphics[width=0.7\hsize]{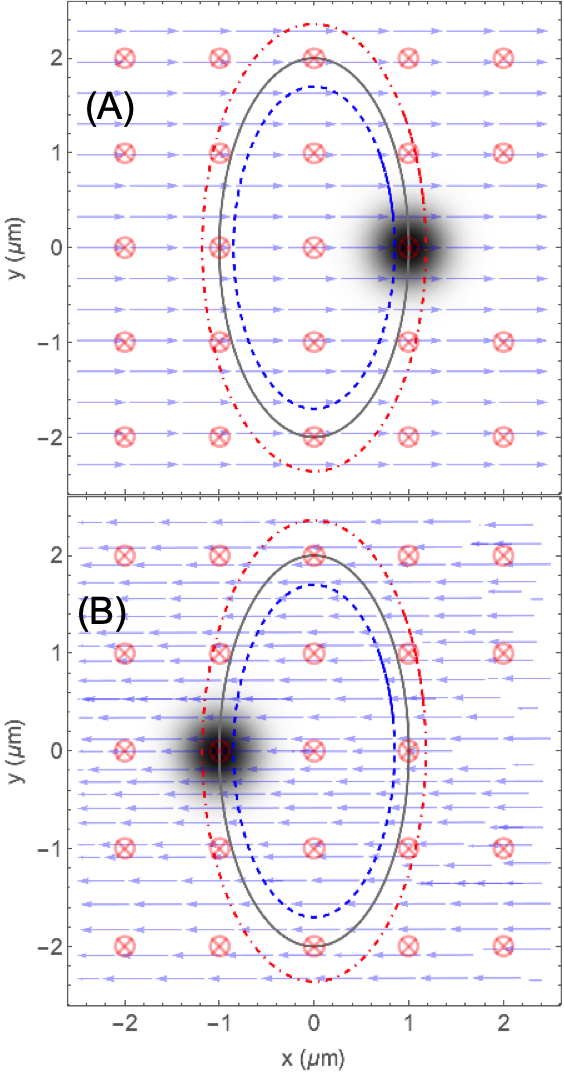}
      \caption{ Dispersionless Motion. Time snapshot of the state evolution (\ref{boostedPsi}) (A) at the beginning of the translation $t=0 ns$ and (B) at $t=6.6 ns$. The black diffused circle represents the electron probability density moving along the full grey curve with frequency $\omega$ without changing its shape.
      The dashed blue and dot-dashed red curves are the trajectories of classical point particles calculated by numerically
      solving the Lorentz force equation with the driving fields given by Eqs. (\ref{RotationB3}), (\ref{RotationE1cl}) and (\ref{RotationE2cl}) of the appendix. 
      This dynamics is achieved by a combination of a rotating electric field
      (blue arrows) given by Eqs. (\ref{RotationE1}) and (\ref{RotationE2}) and a time dependent homogeneous magnetic field $B$ perpendicular to the plane of the figure  (crossed red circles) given by Eq. (\ref{RotationB3}) of the appendix. 
      The values of the parameters are $\epsilon=mc^2$, $B=0.35$T, $a_1=1\mu m$, $a_2=2\mu m$ and $\omega =0.5$ns$^{-1} $.}
  \label{fig:EMF1}
\end{figure}

In Fig.~\ref{fig:EMF1}, the crossed circles represent the time dependent homogeneous 
magnetic field perpendicular to the plane, and the electric field at times 
$\omega t=0$ (Fig. \ref{fig:EMF1} A) and $\omega t=3.3 $ (Fig. \ref{fig:EMF1} B) are displayed as blue arrows in the $x-y$ plane. The dashed blue and dot-dashed
red curves are the trajectories of classical point particles, initially localized at different points within the electron's wavefunction, calculated by numerically
solving the Lorentz force equation with the driving fields given by Eqs. (\ref{RotationB3}), (\ref{RotationE1cl}) and (\ref{RotationE2cl}) of the appendix in
order to show that the derived electromagnetic fields indeed lead to no spreading.
As shown in Sec. C  of the appendix, these electromagnetic fields satisfy 
Maxwell's equations with an electric current but without free charges. The black diffused circle (initially at $x=1 \mu m $ and $y=0$) depicts the Gaussian
state $\psi_b^\dagger\psi_b$ whose shape is preserved 
during its motion along the full grey curve. 

It is important to investigate two different energy regimes of our solutions: The non-relativistic regime $c\gg a_i\omega$ and the
highly relativistic regime $a_i\omega\approx c$ where $a_i$ are the semi-major axis of the elliptical trajectory. Note that if $a_i\omega>c$, $\gamma$ becomes complex. 

The non-relativistic limit $c \to \infty $ of the driving fields given by Eqs. (\ref{RotationE1}), (\ref{RotationE2}) and (\ref{RotationB3}) of the appendix consist of the constant  and homogeneous
magnetic field $B$ along the $z$ direction and the time dependent electric field:
$e\boldsymbol{E}= \omega \left\{ \left(a_2eB-a_1m \omega \right) \cos \omega t, \left(a_1eB-a_2m \omega \right) \sin \omega t,0 \right\}$, $e\boldsymbol{B}=\{0,0,-eB\}$.
This setup can be shown to preserve the Gaussian shape within the Schr\"odinger equation. 
Note that this dynamics can be observed at experimentally 
available values of $B=0.35$T and $|\mathbf{E}| \sim 0.3$V/m employed in Fig.  \ref{fig:EMF1}. 
In such a regime, the radiation energy loss per cycle is infinitesimally (i.e., $11$ orders of magnitude) 
smaller than the electron's kinetic energy. Therefore, the obtained solutions fall within the range of applicability of the Dirac equation. 

The highly relativistic limit $\gamma\gg1$ of the driving fields is given by Eqs. (\ref{EFieldRel}) and (\ref{BFieldRel}) of the appendix.
Contrary to the non-relativistic case, the magnetic field becomes time dependent while the electric field is dependent on both space and time. In the intermediary
regime shown in Fig. \ref{fig:EMF1} the electric field has a weak dependent on both $x$ and $y$ proportional to $\omega^3(a_1^2+a_2^2)eB/(2c^2)$.

\section{A General solution for an electron interacting with the combination of a plane electromagnetic wave with electric and magnetic fields}

The starting point for the construction of the desired solution is the matrix spinor (\ref{MatrixHomo}). The matrix spinor is construct
by applying to  (\ref{MatrixHomo}) the following combination of local Lorentz transformations 
\begin{align}\label{LorentzNull}
\Psi_T=e^{\frac{c\,k\!\!\!/\wedge\mathcal{A}\!\!\!/}{2\omega(p_0-p_z)}}\mathcal{B}_z\Psi e^{\gamma_2\gamma_1\Phi},
\end{align}
where the terms applied to the left of $\Psi$ consist of the following boost along the $z$ direction 
\begin{align}\label{Bz}
\mathcal{B}_z=\mathcal{B}(\mathbf{v}_z),
\end{align}
with $\mathbf{v}_z=\{0,0,v^3(\xi)\}$, $p_0=mcv_0$, $p_z(\xi)=mcv^3(\xi)$ and $\xi=\omega t-\omega z/c$ followed by a combination of boosts and rotations given by
\begin{align}\label{NullBivector}
k\!\!\!/\wedge\mathcal{A}\!\!\!/=f_1'(\xi )\left( \alpha_1+ \boldsymbol{i}\alpha_2\right)+
f_2'(\xi )\left(\alpha_2-\boldsymbol{i}\alpha_1\right)
\end{align}
while the term applied to the right is a rotation about the $\gamma_2\gamma_1$ axis leading to a gauge transformation given by the free function $\Phi$.
The $\omega$ and $k_\mu=\frac{\omega}{c}(1,0,0,1)$ are the laser's frequency and wave vector, respectively. Moreover, the application of the successive
Lorentz transformations $e^{\frac{c\,k\!\!\!/\wedge\mathcal{A}\!\!\!/}{2\omega(p_0-p_z)}}\mathcal{B}_z$ induce the following change of coordinates
\begin{align*}
 x'&= x+\int_0^\xi d\phi\frac{ (p_z(\phi)+p_0)f_1'(\phi)}{m^2\omega ^2 } \\
 y'&= y+\int_0^\xi d\phi\frac{(p_z(\phi)+p_0)f_2'(\phi)}{m^2\omega ^2}.
\end{align*}
The Dirac spinor is then
\begin{align}\label{DiracVolkov}
\psi_T&=\mathcal{N}(\epsilon+mc^2)\sqrt{\frac{p_z(\xi)}{mc}+\frac{p_0}{mc}}e^{-i\Phi-\frac{eB}{4\hbar c}G(x',y')}\times\nonumber\\
   &\left(
\begin{array}{c}
 -\frac{c \left(f_1'(\xi )-i f_2'(\xi )\right) (c m-p_z(\xi
   )+p_0)}{2 \sqrt{2} \omega  \sqrt{c m (c m+p_0)}
   (p_0-p_z(\xi ))} \\
 \frac{\sqrt{c m (c m+p_0)}}{\sqrt{2} c m} \\
 -\frac{c \left(f_1'(\xi )-i f_2'(\xi )\right) (c m-p_z(\xi
   )+p_0)}{2 \sqrt{2} \omega  \sqrt{c m (c m+p_0)}
   (p_0-p_z(\xi ))} \\
 -\frac{p_z(\xi )}{\sqrt{2} \sqrt{c m (c m+p_0)}} \\
\end{array}
\right).
\end{align}

The components of the vector potential given by Eq. (\ref{A-inverted}) are
\begin{align}\label{AlaserandMag}
eA^0&=\frac{\omega  \hbar  \Phi^{(1,0,0)}(\xi ,x,y)}{c}-\frac{(p_z(\xi )+p_0)
   \left(f_1'(\xi )^2+f_2'(\xi )^2\right)}{2 m^2 \omega ^2}\nonumber\\
   &-p_0-\frac{(p_z(\xi )+p_0)}{4m^2c^2\omega} f_1'(\xi )
   eBG^{(0,1)}\left(x',y'\right)\nonumber\\
   &+\frac{(p_z(\xi )+p_0)}{4m^2c^2\omega}f'_2(\xi ) eBG^{(1,0)}\left(x',y'\right),\nonumber\\
eA^1&=\frac{1}{2} \hbar 
   \left(\frac{eB}{2\hbar c}G^{(0,1)}\left(x',y'\right)-2 \Phi ^{(0,1,0)}(\xi ,x,y)\right)\nonumber\\
   &+\frac{c f_1'(\xi )}{\omega },\nonumber\\
eA^2&=-\frac{1}{2} \hbar  \left(\frac{eB}{2\hbar c}G^{(1,0)}\left(x',y'\right)+2 \Phi ^{(0,0,1)}(\xi
   ,x,y)\right)\nonumber\\
   &+\frac{c f_2'(\xi )}{\omega },\nonumber\\
eA^3&=eA^0-p_z(\xi )+p_0.
\end{align}
Note that the free function $\Phi$ can be chosen such that $eA^0=0$. Thus, it amounts to a gauge transformation. 

The corresponding electromagnetic fields are
\begin{align}\label{EMGenVolkov}
e\boldsymbol{B}&=-\frac{eB (p_z(\xi )+p_0) \nabla'^2G\left(x',y'\right)}{4 c^2 m^2 \omega }\left\{f_1'(\xi ),f_2'(\xi ),0\right\}\nonumber\\
   &+\{f_2''(\xi ),-f_1''(\xi ),-\frac{eB}{4c} \nabla'^2G\left(x',y'\right)\},\\
e\boldsymbol{E}&=\left\{ceB_y,-ceB_x,\omega  p_z'(\xi ) \left(1-\frac{p_z(\xi )}{p_0}\right)\right\},\nonumber\\
\end{align}
where $\nabla'^2=\partial^2/\partial x'^2+\partial^2/\partial y'^2$.
The found solutions are written in terms of the free functions $G, f_1, f_2, p_z$ and $\Phi$, being therefore very general.

It is now important to better explain the significance of the local Lorentz transformation $e^{\frac{c\,k\!\!\!/\wedge\mathcal{A}\!\!\!/}{2\omega(p_0-p_z)}}$ and why does it
leads to plane wave fields. First of all, it should be noted that $k\!\!\!/\wedge\mathcal{A}\!\!\!/$ corresponds to a \textit{null bivector}, which means that $(k\!\!\!/\wedge\mathcal{A}\!\!\!/)^2=0$. This property implies that
$$
e^{\frac{c\,k\!\!\!/\wedge\mathcal{A}\!\!\!/}{2\omega(p_0-p_z)}}=\boldsymbol{1}+\frac{c\,k\!\!\!/\wedge\mathcal{A}\!\!\!/}{2\omega(p_0-p_z)}.
$$
Thus, if we make the substitutions $\mathcal{B}_z\Psi\rightarrow U_i$ and 
$$
\Phi=-p_\mu x^\mu/\hbar-\int_0^\xi d\phi\left(\frac{e\mathcal{A}^\mu p_\mu}{k^\mu p_\mu}-\frac{e^2\mathcal{A}^2}{2k^\mu p_\mu}\right),
$$
in Eq. (\ref{LorentzNull}) we recover the well known Volkov states \cite{wolkow1935klasse}. Therefore, replacing $U_i$ in the Volkov states by the more general
local Lorentz transformation $\mathcal{B}_z\Psi$ leads to the addition of extra electromagnetic fields to the plane wave field from the original Volkov spinor. In the
appendix we prove that the spinor (\ref{DiracVolkov}) exactly satisfies the Dirac equation with the vector potential (\ref{AlaserandMag}). It is noteworthy
that the bivector $k\!\!\!/\wedge\mathcal{A}\!\!\!/$ consists of the boost vector $\{f_1'(\xi ),f_2'(\xi ),0\}$ and the rotation vector $\{-f_2'(\xi ),f_1'(\xi ),0\}$ which are mutually orthogonal.
From the expression for the electromagnetic fields (\ref{EMGenVolkov}) we see that in the terms corresponding to the plane wave field inherited from the
Volkov spinor (i.e., terms that don't depend on neither $\nabla'^2G$ nor $p_z(\xi )$), the electric and magnetic field components
are given by $-\frac{d}{d\xi}\{f_1'(\xi ),f_2'(\xi ),0\}$ and $-\frac{d}{d\xi}\{-f_2'(\xi ),f_1'(\xi ),0\}$, respectively. Moreover, the $z$ component of the electric field is a consequence of the local Lorentz boost $\mathcal{B}_z$.

It is instructive to consider what kind of sources will generate the electromagnetic fields (\ref{EMGenVolkov}). From Maxwell's equations, we get
\begin{align}\label{sources}
\rho_e&=\frac{1}{4c\omega}\bigg(\frac{4 c^2 m^2 \omega ^3 p_z'(\xi )^2}{p_0^3}-4 \omega ^3 p_z''(\xi
   ) \left(1-\frac{p_z(\xi )}{p_0}\right)\nonumber\\
  &+ \frac{eB(p_0+p_z(\xi))}{m^2}\left[f_1'(\xi)\frac{\partial}{\partial y'}-f_2'(\xi)\frac{\partial}{\partial x'}\right]\nabla'^2G(x',y')\bigg),\nonumber\\
  \mu_0\boldsymbol{J}&=\bigg\{-\frac{eB}{4c}\frac{\partial}{\partial y'}\nabla'^2G(x',y'),\frac{eB}{4c}\frac{\partial}{\partial x'}\nabla'^2G(x',y'),\frac{\rho_e}{c}\bigg\}.\nonumber\\
\end{align}
Hence, unless $\nabla'^2G(x',y')$ is constant, which happens only if $G(x',y')=c(x'^2+y'^2)$ or if $G(x',y')$ is a harmonic function, the magnetic fields from our solutions will be inhomogeneous.

In what follows, we will consider some particular cases of our general solution (\ref{DiracVolkov}).

\subsection{Solution to the Dirac equation for a particle with a plane electromagnetic wave and a homogeneous magnetic field}
As a rule the free functions of our solutions are chosen such that the source of the given electromagnetic fields (\ref{sources}) have a simple form,
the simplest being source free fields in vacuum.
For instance, by choosing $G(x,y)=c(x^2+y^2)$ and $p_z=0$, the vector potential becomes
\begin{align*}
eA^0&=-\frac{c( f_1'(\xi )^2+ f_2'(\xi )^2)}{2 m \omega ^2}+\frac{\omega  \hbar  \Phi^{(1,0,0)}(\xi ,x,y)}{c}-c m\nonumber\\
   &-\frac{eB \left(y' f_1'(\xi )-x' f_2'(\xi )\right)}{2 m \omega },\nonumber\\
eA^1&=\frac{1}{2} \left(-2 \hbar  \Phi^{(0,1,0)}(\xi ,x,y)+\frac{2 c f_1'(\xi )}{\omega
   }+eB y'\right),\nonumber\\
eA^2&=-\hbar  \Phi^{(0,0,1)}(\xi ,x,y)+\frac{c f_2'(\xi )}{\omega }-\frac{eB
   x'}{2},\nonumber\\
eA^3&=eA^0+mc.
\end{align*}
while the electromagnetic fields are
\begin{align}\label{LaserEM}
e\boldsymbol{B}&=\left\{\frac{-eB f_1'(\xi )}{m \omega }+f_2''(\xi ),\frac{-eB f_2'(\xi )}{m \omega }-f_1''(\xi ),-eB\right\},\nonumber\\
e\boldsymbol{E}&=\left\{ ceB_y,-ceB_x,0\right\}.\nonumber\\
\end{align}
It is straightforward to show that the electromagnetic fields (\ref{LaserEM}) obey the source-free Maxwell's equations, as desired.
This solution is a particular case of the well known Redmond solution \cite{redmond1965solution} provided that $\lim_{\xi\rightarrow-\infty} f_i'(\xi )=\lim_{\xi\rightarrow-\infty} f_i''(\xi )=0$, $i=1,2$
and that the asymptotic spinor corresponds to the ground state of the homogeneous magnetic field.

\subsubsection{Gaussian tracing out a circle without dispersion in the presence of a plane wave field and a homogeneous and constant magnetic field.}
For this example we choose the following functions
\begin{align*}
&f_1(\xi)=a_0(\cos(\xi)-1),\quad f_2(\xi)=-a_0\sin(\xi).
\end{align*}

\begin{figure}
  \centering
      \includegraphics[width=1.0\hsize]{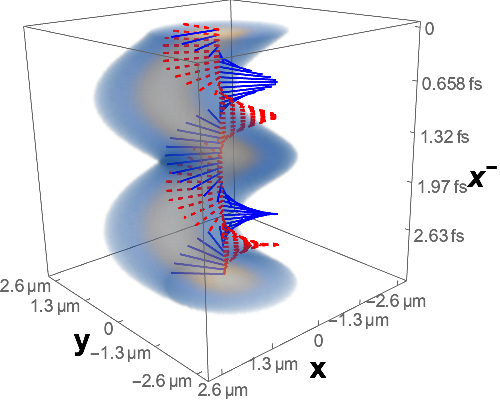}
      \caption{ Dispersionless Motion in a directed plane wave field. The diffuse tube represents the state (\ref{DiracVolkov}).
      This dynamics is achieved by a combination of circularly polarized 800 nm laser field propagating along $x^-=t-z/c$ of intensity $10^{21}W/cm^2$ with electric and magnetic field components represented by the blue full curve and the red dashed curve, respectively and a constant and homogeneous magnetic field $\boldsymbol{B}$ perpendicular to the $x-y$ plane. 
     The values of the parameters are $\epsilon=mc^2$, $a_0=3.24$T, $B=0.13$T and $\omega =eB/m=2.35$fs$^{-1} $.}
  \label{fig:EMF2}
\end{figure}
In Fig.~\ref{fig:EMF2}, the blue full curve represents the electric field while the red dashed curve represents the magnetic field components of a circularly polarized directed plane wave propagating along the $x^-=ct-z$ direction with polarization on the $x-y$ plane. 
It is easy to see that these electromagnetic fields satisfy the source free
Maxwell's equations. The diffuse tube depicts the Gaussian
state  $\psi_T^\dagger\psi_T$ whose shape is preserved 
during its circular motion on the $x-y$ plane. As seen from the explicit expression of the Dirac spinor ( see section D of the appendix) the wavepacket is unbounded along the $z$ axis.

\subsection{Solution to the Dirac equation for a particle in a plane electromagnetic wave and a combination of a homogeneous magnetic field and an electric field}
For this example, we also choose $G(x,y)=c(x^2+y^2)$ but keep $p_z(\xi)$ arbitrary. The electromagnetic fields then become
\begin{align}\label{EMBagrov}
e\boldsymbol{B}&=-\left\{\frac{eB(p_z(\xi)+p_0) f_1'(\xi )}{m^2c\omega},\frac{eB f_2'(\xi ) (p_z(\xi )+p_0)}{c m^2 \omega
   }, eB\right\}\nonumber\\
   &+\{f_2''(\xi ),-f_1''(\xi ),0\},\\
e\boldsymbol{E}&=\left\{ceB_y,-ceB_x,\omega  p_z'(\xi ) \left(1-\frac{p_z(\xi )}{p_0}\right)\right\}.\nonumber\\
\end{align}
Thus, we recover the generalization of the Redmond solution \cite{redmond1965solution} first given by Bagrov et. al. in Ref. \cite{bagrov1975solutions}. 
\comm{A major drawback of Bagrov's solution is that the fields are generally generated by both free charges and currents. For the electromagnetic fields (\ref{EMBagrov}) to obey the source free Maxwell's equations, $p_z(\xi)$ should satisfy the equation
\begin{align}\label{Eqpz}
c^2 m^2 p_z'(\xi)^2-p_0^2 (p_0-p_z(\xi)) p_z''(\xi)=0.
\end{align}
The following function satisfies Eq. (\ref{Eqpz}) 
$$
p_z(\xi)=-\frac{c m \xi  (2 a+\xi )}{2 a (a+\xi )},
$$
where the dimensionless constant $a$ obeys the following conditions: $a>0$ if $\xi>0$ or $a<0$ if $\xi<0$. The corresponding electromagnetic fields are
\begin{align}
e\boldsymbol{B}&=\left\{\frac{a B f_1'(\xi )}{m \omega  (a+\xi )}+f_2''(\xi ),\frac{a B f_2'(\xi )}{m \omega  (a+\xi )}-f_1''(\xi ),B\right\},\\
e\boldsymbol{E}&=\left\{ceB_y,-ceB_x,-\frac{mc\omega}{a}\right\}.
\end{align}
Thus consisting of the combination of a plane electromagnetic wave with antiparallel, homogeneous and constant electric and magnetic fields pointing in the propagation direction of the plane wave field.
}
\section{Outlook.}
We have applied RDI, a new framework for analytically constructing electromagnetic fields 
controlling the dynamics of the Dirac equation, to the case of dispersioneless translation along an \textit{arbitrary} trajectory in the $x-y$ plane.
Illustrations are given for  a Gaussian wavepacket moving along an ellipse and a circle in the $x-y$ plane.

Additionally, 
we found solutions for a Dirac electron driven by the combination of a plane electromagnetic wave with both axial electric and magnetic fields
with non homogeneous perpendicular profiles.
In the process of finding these mew solutions, RDI provided a glimpse of what might come by further exploration of the
full freedom contained in the spinor factorization (\ref{Psi-factorized}). 
Moreover, our illustrations of RDI also hints on a deep connection between the electron motion in external fields described by the Dirac equation and the underlying
geometry of the Lorentz group, the symmetry group of quantum relativistic dynamics

Having illustrated the potential of RDI, the challenges presented upon us are twofold. First, is the task of finding square-integrable solutions
to the Dirac equation for an electron interacting with realistic
laser fields (e.q., gaussian beams). Second is whether or not RDI can be used as means to construct the complete set of eigenvalues and
eigenfunctions for bound state problems. The key to such understanding lies in elucidating the physical and geometrical meanings
of each term in the Dirac and Dirac-Hestenes spinors for the solutions to the Dirac equation. 

\emph{Acknowledgments}
A.G.C. acknowledges financial support from the Humboldt Foundation and insightful discussions with K.Z. Hatsagortryan whose comments greatly improved the presentation of this work.

 \appendix
 
 \begin{widetext}
\section{Proof that $e A\!\!\!/$ satisfy Maxwell's equations}

The importance of finding a real $A\!\!\!/$ through the inversion procedure is the following. First, let us recall the form of Maxwell's equations
\begin{align}\label{Maxwells}
&\rho_e=\vec{\nabla}\cdot\boldsymbol{E},\nonumber\\
&\vec{\nabla}\cdot\boldsymbol{B}=0,\nonumber\\
\mu_0\boldsymbol{J}&=\vec{\nabla}\times\boldsymbol{B}-\frac{\partial}{c^2\partial t}\boldsymbol{E},\nonumber\\
&\vec{\nabla}\times\boldsymbol{E}=-\frac{\partial}{c\partial t}\boldsymbol{B}.
\end{align}
Giving any real function $A_0(t,x,y,z)$ and any real vector field $\boldsymbol{A}=(A_1(t,x,y,z),A_2(t,x,y,z),A_3(t,x,y,z))^T$, set 
\begin{align}
\boldsymbol{B}&=\vec{\nabla}\times \boldsymbol{A},\label{maxwell1}\\
\boldsymbol{E}&=-\vec{\nabla}(A_0)-\frac{\partial}{c\partial t}\boldsymbol{A}\label{maxwell1a}.
\end{align}
By also setting
\begin{align}\label{maxwell2}
\rho_e=\vec{\nabla}\cdot\boldsymbol{E},\quad\mu_0\boldsymbol{J}=\vec{\nabla}\times\boldsymbol{B}-\frac{\partial}{c^2\partial t}\boldsymbol{E},
\end{align}
we get solutions to (\ref{Maxwells}). That this is the case can be easily seem from Eqs. (\ref{maxwell1}) and  (\ref{maxwell1a}) as follows. Since $\boldsymbol{B}$ is the curl of a vector field, then $\vec{\nabla}\cdot\boldsymbol{B}=0$. Moreover, by taking the curl of $\boldsymbol{E}$ we get from (\ref{maxwell1a})
\begin{align}
\vec{\nabla}\times\boldsymbol{E}=-\frac{\partial}{c\partial t}\vec{\nabla}\times\boldsymbol{A}\rightarrow\vec{\nabla}\times\boldsymbol{E}=-\frac{\partial}{c\partial t}\boldsymbol{B}.
\end{align}
Therefore, the vector potential derived from the RDI method is \textit{guaranteed to obey} Maxwell's equations.

\section{Proof that the calculated vector potentials really satisfies the Dirac equation for the given spinors}

We begin by expanding the Dirac equation given in Eq. (\ref{Dirac-Equation}) of the main text
\begin{align}
\label{Dirac-Equation2}
    i c \hbar (\partial_{t}\gamma^0+\partial_x\gamma^1+\partial_y\gamma^2+\partial_z\gamma^3)\psi - (c e A^{0}\gamma_0+ceA^1\gamma_1+ceA^2\gamma_2+ceA^3\gamma_3+mc^2)\psi = 0,  
\end{align}
For the spinor (\ref{boostedPsi}) of the main text, the first term on the right hand side of Eq. (\ref{Dirac-Equation2}) becomes
\begin{align*}
 i c \hbar (\partial_{t}\gamma^0+\partial_x\gamma^1+\partial_y\gamma^2+\partial_z\gamma^3)\psi_b=\bar{\psi},\quad \bar{\psi}=\begin{pmatrix}
        0 \\ \bar{\psi}_2 \\ \bar{\psi}_3 \\ 0
       \end{pmatrix}
 \end{align*}
 where
\begin{align*}
\bar{\psi}_2&=\frac{i \sqrt{\frac{\gamma }{\gamma +1}} \mathcal{N} \left(c^2 m+\epsilon \right)
   e^{-\frac{eB G\left(x',y'\right)}{4 c \hbar }}}{4 \sqrt{2} c^2}\bigg\{\frac{c \left(-eB f'(t) G^{(1,0)}\left(x',y'\right)-eB g'(t)
   G^{(0,1)}\left(x',y'\right)\right)}{\gamma }+\\
   &2 \gamma   \hbar  \left(f'(t) f''(t)+g'(t) g''(t)\right)+i c eB \left(f'(t)
   G^{(0,1)}\left(x',y'\right)-g'(t) G^{(1,0)}\left(x',y'\right)\right)\bigg\},\\
\bar{\psi}_3=&-\frac{\sqrt{\frac{\gamma }{\gamma +1}} \mathcal{N} \left(c^2 m+\epsilon \right)
   e^{-\frac{eB G\left(x',y'\right)}{4 c \hbar }}}{4 \sqrt{2} c^2}\bigg\{\frac{2 c \gamma  eB G^{(0,1)}\left(x',y'\right) \left(c+\frac{\gamma  
   g'(t) \left(-g'(t)-i f'(t)\right)}{c (\gamma +1)}\right)}{\gamma +1}-8 c \gamma  
   \hbar  \left(g''(t)+i f''(t)\right)+\\
   &\frac{2 i c \gamma  eB G^{(1,0)}\left(x',y'\right) \left(c-\frac{\gamma  
   f'(t) \left(f'(t)-i g'(t)\right)}{c (\gamma +1)}\right)}{\gamma +1}+\frac{4 \gamma
   ^2  \hbar  \left(g'(t)+i f'(t)\right) \left((1+2 i) f''(t) g'(t)+f'(t)
   \left(f''(t)-2 i g''(t)\right)\right)}{c (\gamma +1)}\bigg\},\\
   \gamma&=\frac{c}{\sqrt{c^2- f'(t)^2- g'(t)^2}}
\end{align*}
It is then straightforward to show that upon substituting the vector potential given in Eq. (\ref{VecPot}) of the main text on the second term on the right hand side of (\ref{Dirac-Equation2}) we get
\begin{align*}
(c e A^0\gamma_0+ceA^1\gamma_1+ceA^2\gamma_2+ceA^3\gamma_3+mc^2)\psi_b=\bar{\psi}.
\end{align*}
Thus, the Dirac equation (\ref{Dirac-Equation2}) is exactly satisfied.

For the spinor (\ref{DiracVolkov}) of the main text, the first term on the right hand side of Eq. (\ref{Dirac-Equation2}) becomes
\begin{align*}
 i c \hbar (\partial_{t}\gamma^0+\partial_x\gamma^1+\partial_y\gamma^2+\partial_z\gamma^3)\psi_T=\phi,\quad \phi=\begin{pmatrix}
        \phi_1 \\ \phi_2 \\ \phi_3 \\ \phi_4
       \end{pmatrix}
 \end{align*}
where
\begin{align*}
\phi_1&=-\sqrt{\rho}e^{-i\Phi}\frac{c \hbar  p_z(\xi )}{2
   \sqrt{2} \sqrt{c m (c m+p_0)}}\bigg\{ \left(\frac{eB G^{(0,1)}\left(x',y'\right)}{2 c
   \hbar }+\frac{i eB G^{(1,0)}\left(x',y'\right)}{2 c \hbar }\right)+i \left(2 \Phi^{(0,0,1)}(\xi ,x,y)+2 i \Phi^{(0,1,0)}(\xi ,x,y)\right)\bigg\},\\
   \phi_2&=\frac{(c m+p_0-p_z(\xi ))\sqrt{\rho}e^{-i\Phi}}{4 \sqrt{2} \omega  \sqrt{c m (c
   m+p_0)} (p_0-p_z(\xi ))}\bigg\{ \left(\frac{1}{2} i c eB
   \left(f_1'(\xi )+i f_2'(\xi )\right)
   G^{(1,0)}\left(x',y'\right)+\frac{1}{2} c \left(f_1'(\xi )+i f_2'(\xi
   )\right) eBG^{(0,1)}\left(x',y'\right)\right)\\
   &+2 \left(c^2 \hbar  \left(f_2'(\xi )+i f_1'(\xi )\right) \Phi
   ^{(0,0,1)}(\xi ,x,y)+c^2 \hbar  \left(f_1'(\xi )-i f_2'(\xi )\right)
   \Phi ^{(0,1,0)}(\xi ,x,y)-2 \omega ^2 \hbar  (p_0-p_z(\xi )) \Phi
   ^{(1,0,0)}(\xi ,x,y)\right)\bigg\},\\
   \phi_3&=-\sqrt{\rho}e^{-i\Phi}\frac{ \hbar  \sqrt{mc(mc+p_0)}}{2
   \sqrt{2} m}\bigg\{ \left(\frac{eB G^{(0,1)}\left(x',y'\right)}{2 c
   \hbar }+\frac{i eB G^{(1,0)}\left(x',y'\right)}{2 c \hbar }\right)+i \left(2 \Phi^{(0,0,1)}(\xi ,x,y)+2 i \Phi^{(0,1,0)}(\xi ,x,y)\right)\bigg\},\\
   \phi_4&=\frac{(c m+p_0-p_z(\xi ))\sqrt{\rho}e^{-i\Phi}}{4 \sqrt{2} \omega  \sqrt{c m (c
   m+p_0)} (p_0-p_z(\xi ))}\bigg\{\frac{1}{2} c eB \left(f_2'(\xi )-i f_1'(\xi )\right)
   G^{(1,0)}\left(x',y'\right)-\frac{1}{2} c eB \left(f_1'(\xi )+i
   f_2'(\xi )\right) G^{(0,1)}\left(x',y'\right)\\
   &+2 \Big[-i c^2 \hbar  \left(f_1'(\xi )-i f_2'(\xi )\right) \Phi
   ^{(0,0,1)}(\xi ,x,y)-c^2 \hbar  \left(f_1'(\xi )-i f_2'(\xi )\right)
   \Phi ^{(0,1,0)}(\xi ,x,y)\\
   &+2 \omega ^2 \hbar  \left(\sqrt{c^2 m^2+p_z(\xi
   )^2}-p_z(\xi )\right) \Phi ^{(1,0,0)}(\xi ,x,y)\Big] \bigg\},\\
   \sqrt{\rho}&=\mathcal{N}(\epsilon+mc^2)\sqrt{\frac{p_z(\xi)}{mc}+\frac{p_0}{mc}}e^{-\frac{eB}{4\hbar c}G(x',y')}.
 \end{align*}
 It is then straightforward to show that upon substituting the vector potential given in Eq. (\ref{AlaserandMag}) of the main text on the second term on the right hand side of (\ref{Dirac-Equation2}) we get
\begin{align*}
(c e A^0\gamma_0+ceA^1\gamma_1+ceA^2\gamma_2+ceA^3\gamma_3+mc^2)\psi_T=\phi.
\end{align*}
Thus, the Dirac equation (\ref{Dirac-Equation2}) is exactly satisfied.
\section{Dispersionless motion along an elliptical path: Gaussian state in 2D}\label{Sec:Rotation2D}

For this particular case the Dirac spinor $\psi_b$ is
\begin{align}\label{DiracSpinor}
 \psi_b = \left(
\begin{array}{c}
0\\
 \frac{\mathcal{N} c m \sqrt[4]{c^2- \omega ^2 \left(a_1^2 \sin ^2(t \omega
   )+a_2^2 \cos ^2(t \omega )\right)} \sqrt{\frac{c^2}{\sqrt{c^2- \omega ^2
   \left(a_1^2 \sin ^2(t \omega )+a_2^2 \cos ^2(t \omega )\right)}}+c}
   \exp \left(-\frac{B \left((x-a_1  \cos (t \omega ))^2+(y-a_2  \sin (t
   \omega ))^2\right)}{4 \hbar }\right)}{\sqrt{2}} \\
 \frac{\mathcal{N} c^2 m  \omega  (-a_1 \sin (t \omega )+i a_2 \cos (t
   \omega )) \exp \left(-\frac{B \left((x-a_1  \cos (t \omega ))^2+(y-a_2
    \sin (t \omega ))^2\right)}{4 \hbar }\right)}{\sqrt{2} \sqrt[4]{c^2- \omega ^2
   \left(a_1^2 \sin ^2(t \omega )+a_2^2 \cos ^2(t \omega )\right)}
   \sqrt{\frac{c^2}{\sqrt{c^2- \omega ^2 \left(a_1^2 \sin ^2(t \omega
   )+a_2^2 \cos ^2(t \omega )\right)}}+c}} \\
   0\\
\end{array}
\right).
\end{align}
The corresponding electric field is 
\begin{align}
 eE_1 &=\frac{\cos (t \omega )}{2c^2\gamma} \left(\frac{\omega ^2 \left(c^2-a_2^2 \omega ^2\right)
   \left(a_2 \omega  \left(a_1^2 eB-\hbar \right)-2 a_1 c^2
   m\right)}{c^2\gamma ^2}+a_2 eB \gamma ^2 \omega  \left(\frac{a_2^2 \omega
   ^2}{\gamma ^2}+2c^2\right)\right)\nonumber\\
   &-\frac{a_1 a_2 eB x \omega  \left(c^2\gamma ^2-\frac{\left(c^2-a_1^2
   \omega ^2\right) \left(c^2-a_2^2 \omega ^2\right)}{c^2\gamma
   ^2}\right)}{2c^2\gamma(a_1^2-a_2^2)}-\frac{1}{4c^2\gamma} eB y \omega ^3 \sin (2 t \omega ) \left(\frac{a_1^2
   \left(c^2-a_2^2 \omega ^2\right)}{c^2\gamma ^2}+a_2^2\right), \label{RotationE1}\\
 eE_2 &=\frac{\sin (t \omega )}{2c^2\gamma} \left(a_1 eB \omega  \left(a_1^2 \omega
   ^2+2 c^2\gamma ^2\right)-\frac{\omega ^2 \left(c^2-a_1^2 \omega
   ^2\right) \left(2 a_2 c^2 m-a_1 \omega 
   \left(a_2^2 eB-\hbar \right)\right)}{c^2\gamma ^2}\right)\nonumber\\
   &+\frac{a_1 a_2 eB y \omega  \left(c^2\gamma ^2-\frac{\left(c^2-a_1^2
   \omega ^2\right) \left(c^2-a_2^2 \omega ^2\right)}{\gamma
   ^2c^2}\right)}{2c^2\gamma(a_1^2-a_2^2)}-\frac{1}{4c^2\gamma} eB x \omega ^3 \sin (2 t \omega ) \left(\frac{a_2^2
   \left(c^2-a_1^2 \omega ^2\right)}{c^2\gamma ^2}+a_1^2\right),\label{RotationE2}\\
 eE_3 &= 0. \label{RotationE3}
\end{align}
The magnetic field is 
\begin{align}
 eB_1 =& 0, \label{RotationB1} \\
 eB_2 =& 0,\\
 eB_3 =&  -\frac{eB \left(2 c^2- \omega ^2 \left(a_1^2 \sin ^2(t \omega )+a_2^2
   \cos ^2(t \omega )\right)\right)}{2 c \sqrt{c^2- \omega ^2 \left(a_1^2
   \sin ^2(t \omega )+a_2^2 \cos ^2(t \omega )\right)}}=-\frac{eB}{2}\left(\gamma+\frac{1}{\gamma}\right). \label{RotationB3}
 \end{align}


The obtained electromagnetic field obeys Maxwell's equations 
\begin{align}
 \nabla \cdot \mathbf{E} &= 0, \label{MaxwellEq1} \\
 \nabla \cdot \mathbf{B} &= 0, \\
 \nabla \times \mathbf{E} + \frac{\partial }{\partial t} \mathbf{B} &= 0, \\
 \nabla \times \mathbf{B} - \frac{1}{c^2} \frac{\partial}{\partial t} \mathbf{E} &=  \mu_0 \mathbf{J}, \label{MaxwellEq4}
\end{align}
with the current $\mathbf{J}$ 
\begin{align}
 \mu_0 e J^1 =& \label{RotationMaxwellJ1}
-\frac{\partial}{c^2\partial t}eE_1,
 \\
   \mu_0 e J^2 =&
-\frac{\partial}{c^2\partial t}eE_2, \\
\mu_0 e J^3 =& 0. \label{RotationMaxwellJ3}
\end{align}

The non-relativistic limit of the electromagnetic field (\ref{RotationE1})-(\ref{RotationB3}) is 
\begin{align}
 e\mathbf{E}_{nr} &= 
  \left\{ \omega  \cos (t \omega ) \left(eBa_2-a_1m \omega \right),\omega  \sin (t \omega ) \left(eBa_1-a_2m \omega \right), 0\right\}, \\
 e\mathbf{B}_{nr} &= \{ 0,0,-eB \}.
\end{align}
It is noteworthy that the time dependence of the magnetic field as well as the space dependence of the electric field are more pronounced in the high energy regime, in which $\omega a_i\approx c$, where $a_i$ is the semi-major axis of the ellipse. In this regime $\gamma\gg1$ and the electric and magnetic fields becomes
\begin{align}
e\boldsymbol{B}_r&\approx-\left\{0,0,\frac{\gamma eB}{2}\right\},\label{BFieldRel}\\
e\boldsymbol{E}_r&\approx\gamma eB\omega\left\{a_2\cos\omega t-\frac{a_2a_1x}{2(a_1^2-a_2^2)},a_1\sin\omega t+\frac{a_1a_2y}{2(a_1^2-a_2^2)},0\right\}\label{EFieldRel}.
\end{align}

In the classical limit $\hbar\rightarrow0$, the magnetic field (\ref{RotationB3}) remains unchanged while the electric field becomes
\begin{align}
 e\tilde{E}_1 &=\frac{\cos (t \omega )}{2c^2\gamma} \left(\frac{\omega ^2 \left(c^2-a_2^2 \omega ^2\right)
   \left(a_2 \omega a_1^2 eB-2 a_1 c^2
   m\right)}{c^2\gamma ^2}+a_2 eB \gamma ^2 \omega  \left(\frac{a_2^2 \omega
   ^2}{\gamma ^2}+2c^2\right)\right)\nonumber\\
   &-\frac{a_1 a_2 eB x \omega  \left(c^2\gamma ^2-\frac{\left(c^2-a_1^2
   \omega ^2\right) \left(c^2-a_2^2 \omega ^2\right)}{c^2\gamma
   ^2}\right)}{2c^2\gamma(a_1^2-a_2^2)}-\frac{1}{4c^2\gamma} eB y \omega ^3 \sin (2 t \omega ) \left(\frac{a_1^2
   \left(c^2-a_2^2 \omega ^2\right)}{c^2\gamma ^2}+a_2^2\right), \label{RotationE1cl}\\
 e\tilde{E}_2 &=\frac{\sin (t \omega )}{2c^2\gamma} \left(a_1 eB \omega  \left(a_1^2 \omega
   ^2+2 c^2\gamma ^2\right)-\frac{\omega ^2 \left(c^2-a_1^2 \omega
   ^2\right) \left(2 a_2 c^2 m-a_1 \omega a_2^2 eB\right)}{c^2\gamma ^2}\right)\nonumber\\
   &+\frac{a_1 a_2 eB y \omega  \left(c^2\gamma ^2-\frac{\left(c^2-a_1^2
   \omega ^2\right) \left(c^2-a_2^2 \omega ^2\right)}{\gamma
   ^2c^2}\right)}{2c^2\gamma(a_1^2-a_2^2)}-\frac{1}{4c^2\gamma} eB x \omega ^3 \sin (2 t \omega ) \left(\frac{a_2^2
   \left(c^2-a_1^2 \omega ^2\right)}{c^2\gamma ^2}+a_1^2\right),\label{RotationE2cl}\\
\end{align}

The current constructed from the Dirac spinors is $J_D = \Psi \gamma_0\tilde{\Psi}$, whose components $J_D^{\mu} = Tr(\Psi\gamma_0\tilde{\Psi} \gamma_{\mu})/4$ are
\begin{align}
 J_D^0 =&\mathcal{N}^2 c^4 m^2 \exp \left(-\frac{B \left((x-a_1  \cos (t \omega
   ))^2+(y-a_2  \sin (t \omega ))^2\right)}{2 \hbar }\right), \label{EqJD1} \\
J_D^1 =&- \mathcal{N}^2 a_1 c^3 m^2  \omega  \sin (t \omega ) \exp \left(-\frac{B
   \left((x-a_1  \cos (t \omega ))^2+(y-a_2  \sin (t \omega ))^2\right)}{2
   \hbar }\right),\\
J_D^2 =& \mathcal{N}^2 a_2 c^3 m^2  \omega  \cos (t \omega ) \exp \left(-\frac{B
   \left((x-a_1  \cos (t \omega ))^2+(y-a_2  \sin (t \omega ))^2\right)}{2
   \hbar }\right),\\
 J_D^3 =& 0. \label{EqJD4}
\end{align}

The velocity associated with the current $J_D$ is obtained as $v^{k} = c J_D^{k}/ \rho$
\begin{align}
  v^1 =& - a_1\omega \sin(\omega t), \\
  v^2 =&   a_2\omega \cos(\omega t),
\end{align}
with the magnitude given by $|\mathbf{v}| =  \omega\sqrt{a_1^2\cos(\omega t)^2+a_2^2\sin(\omega t)^2}$. 
Thus, superluminal propagation is avoided if $a_i\omega <c$, $i=1,2$. 

The current $\mathbf{J}_e$ (\ref{RotationMaxwellJ1})-(\ref{RotationMaxwellJ3}) entering Maxwell's equations (\ref{MaxwellEq1})-(\ref{MaxwellEq4}) is related to the current $\mathbf{J}_D$ (\ref{EqJD1})-(\ref{EqJD4})  constructed from the Dirac spinor in the following way: The Maxwell current $\mathbf{J}_e$ creates the electromagnetic field (\ref{RotationE1})-(\ref{RotationB3}) steering the Dirac wave packet (\ref{DiracSpinor}). Moving along an elliptical trajectory, a Dirac electron yields the current $\mathbf{J}_D$ emitting radiation. If the radiation losses are large, the proposed solutions may not work. Therefore, the calculated electromagnetic fields are physically meaningful if the electron kinetic energy is much larger than the energy emitted via radiation. The dispersionless rotation shown in Fig. 1 of the main text obeys well this criterion because the radiative energy loss per period is $\propto 10^{-32}$J, whereas the electron kinetic energy is $\propto 10^{-21}$J.

\section{Gaussian tracing out a circle without dispersion in the polarization plane of a circularly polarized directed plane wave field}
For this particular case the Dirac spinor $\psi_T$ is
\begin{align}
\psi_T	=e^{-i\Phi-\frac{eB \left(\left(a_0 c (\cos \left(\xi\right)-1)+m x \omega ^2\right)^2+\left(-a_0 c \sin
   \left(\xi\right)+m y \omega ^2\right)^2\right)}{4
   m^2 \omega ^4 \hbar }}\left(
\begin{array}{c}
 \frac{a_0 (\sin (\xi )-i \cos (\xi ))}{2 m \omega } \\
 1 \\
 \frac{a_0 (\sin (\xi )-i \cos (\xi ))}{2 m \omega } \\
 0 \\
\end{array}
\right)
\end{align}
with
$$
\Phi=\frac{c \left(c \xi  \left(a_0^2 (eB+m \omega )+2 m^3 \omega
   ^3\right)+a_0 eB \left(m \omega ^2 (x \sin (\xi )+y \cos (\xi
   ))-a_0 c \sin (\xi )\right)\right)}{2 m^2 \omega ^4 \hbar }.
 $$
 
 The current constructed from the Dirac spinors is $J_D = \Psi_T \gamma_0\tilde{\Psi}_T$, whose components $J_D^{\mu} = Tr(\Psi_T\gamma_0\tilde{\Psi}_T \gamma_{\mu})/4$ are
\begin{align}
 J_D^0 =&e^{-\frac{eB \left(\left(a_0 c (\cos \left(\xi\right)-1)+m x \omega ^2\right)^2+\left(-a_0 c \sin
   \left(\xi\right)+m y \omega ^2\right)^2\right)}{2
   m^2 \omega ^4 \hbar }}\left(1+\frac{a_0^2}{2 m^2
   \omega ^2}\right),  \\
J_D^1 =&e^{-\frac{eB \left(\left(a_0 c (\cos \left(\xi\right)-1)+m x \omega ^2\right)^2+\left(-a_0 c \sin
   \left(\xi\right)+m y \omega ^2\right)^2\right)}{2
   m^2 \omega ^4 \hbar }}\left(\frac{a_0 \sin \left(\xi\right)}{m \omega }\right),\\
J_D^2 =&e^{-\frac{eB \left(\left(a_0 c (\cos \left(\xi\right)-1)+m x \omega ^2\right)^2+\left(-a_0 c \sin
   \left(\xi\right)+m y \omega ^2\right)^2\right)}{2
   m^2 \omega ^4 \hbar }}\left(\frac{a_0 \cos \left(\xi\right)}{m \omega }\right),\\
 J_D^3 = &e^{-\frac{eB \left(\left(a_0 c (\cos \left(\xi\right)-1)+m x \omega ^2\right)^2+\left(-a_0 c \sin
   \left(\xi\right)+m y \omega ^2\right)^2\right)}{2
   m^2 \omega ^4 \hbar }}\left(\frac{a_0^2}{2 m^2
   \omega ^2}\right). 
\end{align}
 The spin density constructed from the Dirac spinors is $\rho s = \Psi_T \gamma_3\tilde{\Psi}_T$, whose components $\rho s^{\mu} = Tr(\Psi_T\gamma_3\tilde{\Psi}_T \gamma_{\mu})/4$ are
\begin{align}
 \rho s^0 =&e^{-\frac{eB \left(\left(a_0 c (\cos \left(\xi\right)-1)+m x \omega ^2\right)^2+\left(-a_0 c \sin
   \left(\xi\right)+m y \omega ^2\right)^2\right)}{2
   m^2 \omega ^4 \hbar }}\left(\frac{a_0^2}{2 m^2
   \omega ^2}\right),  \\
\rho s^1 =&e^{-\frac{eB \left(\left(a_0 c (\cos \left(\xi\right)-1)+m x \omega ^2\right)^2+\left(-a_0 c \sin
   \left(\xi\right)+m y \omega ^2\right)^2\right)}{2
   m^2 \omega ^4 \hbar }}\left(\frac{a_0 \sin \left(\xi\right)}{m \omega }\right),\\
\rho s^2 =&e^{-\frac{eB \left(\left(a_0 c (\cos \left(\xi\right)-1)+m x \omega ^2\right)^2+\left(-a_0 c \sin
   \left(\xi\right)+m y \omega ^2\right)^2\right)}{2
   m^2 \omega ^4 \hbar }}\left(\frac{a_0 \cos \left(\xi\right)}{m \omega }\right),\\
 \rho s^3 = &e^{-\frac{eB \left(\left(a_0 c (\cos \left(\xi\right)-1)+m x \omega ^2\right)^2+\left(-a_0 c \sin
   \left(\xi\right)+m y \omega ^2\right)^2\right)}{2
   m^2 \omega ^4 \hbar }}\left(-1+\frac{a_0^2}{2 m^2
   \omega ^2}\right). 
\end{align}

\end{widetext}
\bibliography{bib-proposal}

\end{document}